\documentclass[prb,floatfix,reprint,superscriptaddress]{revtex4-1} 

\usepackage{graphicx}
\usepackage{textcomp}
\usepackage[english]{babel}
\usepackage[utf8]{inputenc}
\usepackage[T1]{fontenc}
\usepackage{hyperref}
\usepackage{amsmath}
\usepackage{amssymb}
\usepackage[amssymb]{SIunits}
\usepackage{physics}
\usepackage{bm}

\usepackage{todonotes}

\newcommand{\UGAAddress}{Univ. Grenoble Alpes, F-38000 Grenoble, France}
\newcommand{\CNRSAddress}{CNRS, Inst. NEEL, "Nanophysique et semiconducteurs" group, F-38000 Grenoble, France}
\newcommand{\CEAAddress}{CEA, INAC, "Nanophysique et semiconducteurs" group, F-38000 Grenoble, France}

\begin{document} 
\title[Light-Hole Exciton]{Light-hole Exciton in Nanowire Quantum Dot}

\author{Mathieu Jeannin}

\author{Alberto Artioli}
\affiliation{\UGAAddress}
\affiliation{\CNRSAddress}

\author{Pamela Rueda-Fonseca}
\affiliation{\UGAAddress}
\affiliation{\CNRSAddress}
\affiliation{\CEAAddress}

\author{Edith Bellet-Amalric}

\author{Kuntheak Kheng}
\affiliation{\UGAAddress}
\affiliation{\CEAAddress}

\author{Régis André}

\author{Serge Tatarenko}

\author{Joël Cibert}

\author{David Ferrand}

\author{Gilles Nogues}
\affiliation{\UGAAddress}
\affiliation{\CNRSAddress}
\email{gilles.nogues@neel.cnrs.fr}


\begin{abstract}
Quantum dots inserted inside semiconductor nanowires are
	extremely promising candidates as building blocks for solid-state based quantum
	computation and communication. They provide very high crystalline and optical
	properties and offer a convenient geometry for electrical contacting. 
	Having a complete determination and full control of their emission
	properties is one of the key goals of nanoscience researchers. 
	Here we use strain as a tool to create in a single magnetic
	nanowire quantum dot a light-hole exciton, an optically active
	quasiparticle formed from a single electron bound to a single light hole. 
	In this frame, we provide a general description of the mixing within the hole quadruplet 
	induced by strain or confinement. 
	A multi-instrumental combination of cathodoluminescence, polarisation-resolved
	Fourier imaging and magneto-optical spectroscopy, allow us to fully characterize
	the hole ground state, including its valence band mixing with heavy hole states.
\end{abstract}

\maketitle 

\section{Introduction}
Semiconductor quantum dots are seen as important elements for integrated quantum
simulation and communication.\cite{Michler2000,Zrenner2002,Shields2007} They
can act as static qubits, encoding information either onto their orbital or spin
state. They can also serve as a deterministic source of flying qubits using
single\cite{Santori2001,claudon2010highly} or entangled
photons.\cite{Santori2002,Akopian2006,DousseSenellart_Ultrabrightsourceentangled_10} 
In this perspective, hole spins are particularly interesting because of their weak 
hyperfine coupling to surrounding spin bath compared to electrons.\cite{Bulaev2005,Heiss2007} 
Furthermore working with light-holes pave the way to 
new information technology protocols like direct manipulation of the hole spin 
state with RF fields,\cite{SleiterBrinkman_Usingholesin_06} efficient control 
of a magnetic impurity spin coupled to a quantum 
dot,\cite{ReiterAxt_Coherentcontrolsingle_11} or spin state tomography of the 
electron inside the dot.\cite{KosakaEdamatsu_Spinstatetomography_09} However 
most of the previous studies have concerned so far heavy-hole states, 
because they are energetically favored for a majority of quantum dot 
heterostructures for which confinement and strain lift the degeneracy of 
the valence band.\cite{Bester2003,He2004} Hence a way to address light-holes  
is to promote them as the valence band ground state by engineering 
the strain inside the dot. For epitaxially grown dots this requires 
technologically intensive methods, such as the fabrication of deformable 
membranes containing the dots.\cite{Huo2013} Another very promising strategy 
is to embed the dot inside a nanowire.\cite{Niquet2008} This bottom-up 
approach produces high quality heterostructures. 
It offers a way to control both the carriers confinement through the 
geometrical shape of the dot, and its internal strain by adding a shell 
of a different material around the nanowire core.\cite{Zielinski2013,Ferrand2014} 
To be short, in most nanostructures the low-gap material has a larger lattice parameter. 
In a flat quantum dot as resulting from Stranski-Krastanow growth, it is well known that 
the hole ground state has a main heavy-hole character. 
The most frequent case is that of InAs dots in GaAs, but this is true also for 
CdTe dots in ZnTe.\cite{Lafuente-Sampietro2015} 
This is due to the stronger effect of confinement along the growth axis, 
and to the strongest component of the mismatch strain which is compressive in the plane. 
In a core-shell nanowire made of the same materials, the confinement is stronger in the plane, 
and the strongest component of the mismatch strain is compressive along the axis. 
As a result, both confinement and mismatch strain conspire to make the ground state 
a light-hole state.\cite{Ferrand2014} When increasing the height of a quantum dot in a nanowire, 
a crossing is expected, from the heavy-hole ground state in a flat quantum dot to 
a light-hole ground state in an elongated quantum dot.\cite{Zielinski2013} 
Note that nanowire structures are very flexible, and a heavy-hole ground state can be found 
also in a core-shell nanowire if the lattice mismatch induces 
a tensile strain in the core.\cite{Ferrand2014} 
An additional important property of a semiconductor nanowire is that it acts as a dielectric antenna, 
modulating the coupling of the different exciton transitions to light modes 
which changes their radiation pattern.
\cite{Ruda2006,claudon2010highly,BleuseLalanne_InhibitionEnhancementand_11,Grzela2012, Bulgarini2014}

Here we provide a complete study of a light-hole quantum dot (QD) in a
core-shell nanowire. The dot contains a large fraction of magnetic dopants
in order to enhance the Zeeman shift for spintronics applications. This prevents 
a direct spectroscopic evidence of its light-hole character by measuring its 
fine structure.\cite{Huo2013} We show nevertheless that the detailed observation of the
polarisation state of the QD far field radiation pattern is enough to prove its
light-hole nature and provides a wealth of information about its mixing with
heavy-hole states. Our results are in agreement with numerical simulations of the
QD emission within the full nanowire structure. It is confirmed by studying the giant
Zeeman shift and the polarisation of the excitonic transition under an external magnetic field. 
The magneto-optical spectroscopy reveals a heavy-hole excited state at high field, 
thus providing an order of magnitude for the valence band splitting. 
Our method is simple, and requires no extra processing of the 
sample\cite{Tonin2012,Huo2013} as far as the nanowire is
isolated from its neighbours.

\section{Light-hole and heavy-hole properties and anisotropy}

\begin{figure}[!ht]
\centering
\includegraphics[width=\linewidth]{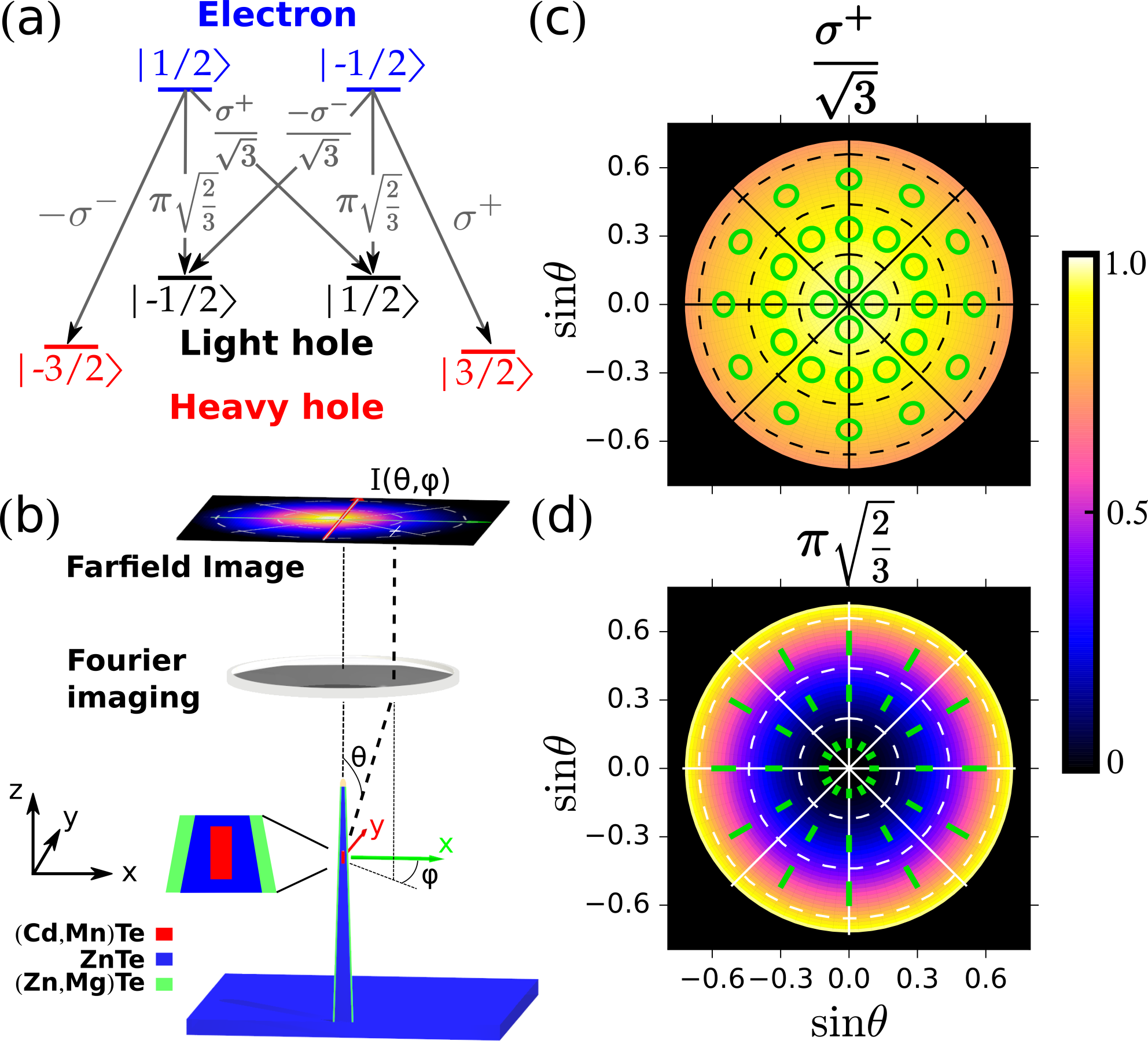}
\caption{
    		(a), Energy levels inside the QD participating to the
	    luminescence, along with the corresponding transitions with their oscillator
	    strength and polarisation. 
	    (b), Schematic of the experiment. The light emission is mapped onto the Fourier plane to
	    allow polarisation analysis of the photons with respect to the direction $(\theta,\varphi)$. 
	    The inset present a zoom of the nanowire in the vicinity of the quantum dot.   
	    (c,d), Theoretical far field intensity maps for the 
	    $\sigma$ or $\pi$ transition, respectively. 
	    The QD is assumed to be in an infinite space made of ZnTe. 
	    The green lines in (c) and (d) represent the time evolution the electric 
	    field in the $(x,y)$ plane for a set of directions. For each direction the electric field origin is centered on the $(\theta,\varphi)$ coordinate.
	    }
\label{fig:Theory}
\end{figure}

The distinction between light- and heavy-hole arises
when the top of the valence band fourfold degeneracy
of zinc-blende semiconductors is lifted by, for instance, strain or
confinement. Generally, a relevant axis of symmetry $z$ appears, such as
the growth axis for self-assembled quantum dots or a
nanowire. Eigenstates are then Kramers doublets, characterized by
the projection of their total spin onto $z$
for the electron ($\pm 1/2$), the light-hole (LH, $\pm 1/2$) and the
heavy-hole  (HH, $\pm 3/2$). The HH has a strong magnetic
anisotropy, with its spin $\pm 3/2$ along $z$ but a vanishing
Land\'{e} factor in the normal $xy$-plane. It exhibits also a strong
optical anisotropy, with dipolar electric transitions matrix
elements towards electron states in the $xy$-plane 
(called $\sigma$ transitions hereafter), see Fig.~\ref{fig:Theory}(a). 
By contrast, the LH has a finite Land\'{e} factor in the $xy$-plane 
and a smaller one along $z$. Optically, it 
presents both $\sigma$ and $\pi$ polarised transition matrix elements.


We use the hole formalism to describe the top of the valence band: 
this is more convenient if one has in mind the optical manipulation 
of holes in a quantum dot \cite{Huo2013} or carrier induced magnetic interactions 
in a dilute magnetic semiconductor.\cite{Dietl1997} As a result, the hole ground state 
is at lower energy, and the light-hole/heavy-hole splitting $\Delta_{LH}$ 
is negative if the ground state is a light hole. 
$\Delta_{LH}<0$ implies a LH ground state. However, in real life quantum
dot, confinement potential and strain (uniform and inhomogeneous) create
additional anisotropy components which break the circular symmetry around $z$
and hence mix the light- and heavy-hole states. This mixing is usually described
by two additional complex numbers $\sigma e^{-i  \chi}$ and 
$\rho e^{-2 i\psi}$.\cite{BirPikus} In appendix~\ref{app:theo}, we
show in details that the resulting $4\times4$ Hamiltonian, with one real number
$\Delta_{LH}$ and two complex numbers $\rho e^{-2i\psi}$ and $\sigma
e^{-i\chi}$, is the most general spin Hamiltonian describing an
isolated spin quadruplet and respecting time reversal symmetry. 
All three terms have various origins, including uniform and inhomogeneous 
strain, and confinement. 
As a consequence of the mixing the true QD hole eigenstates are
linear combinations of the pure heavy- and light-hole states defined by $z$, and
the resulting dipole transitions and spin properties are changed accordingly.

The spin Hamiltonian takes a much simpler form in the frame $(x_0,y_0,z_0)$ 
which diagonalizes the anisotropy tensor (see appendix~\ref{app:theo}). 
It then depends only on two real parameters: 
$\Delta_{LH0}$ which describes the LH/HH
splitting along the principal anisotropy axis $z_0$ and
$\rho_0$ which describes the transverse anisotropy in the $(x_0, y_0)$ plane.
Three Euler angles $\hat{\alpha}$, $\hat{\beta}$ and $\hat{\gamma}$ 
are necessary to characterize the transformation from the laboratory 
frame to the anisotropy frame. 
If $\hat{\beta} \ll 1$, $\hat{\beta}$ and $\hat{\gamma}$ are 
the spherical coordinates $(\theta,\varphi)$ of axis $z_0$. 
$\hat{\alpha}+\hat{\gamma}$ characterizes the
direction of transverse anisotropy. 
The parameter $\sigma$ measured in the laboratory axes 
does not really represent a mixing, but the result of the 
tilt $\hat{\beta}$ between $z$ and $z_0$.
We stress this physical
interpretation of the two "mixing parameters" measured in the
laboratory frame, because it has direct practical consequences:
$\rho e^{-2i\psi}$ describes a real mixing, which
requires that a shear strain be applied in order to compensate for it.\cite{Huo2013} 
On the other hand the effect of $\sigma$ (which can also be due 
to strain and confinement anisotropy) can be compensated by an appropriate 
tilt of the optical axis.

In our setup we collect the nanowire light by placing a
microscope objective (numerical aperture NA=0.72) on the $z$ axis
[Fig.~\ref{fig:Theory}(b)]. A set of additional optical elements
allows to image onto a CCD camera the intensity $I(\theta,\varphi)$
emitted in a direction $(\theta,\varphi)$.\cite{Grzela2012,Bulgarini2014} 
Figures \ref{fig:Theory}(c-d) represent the theoretical colormaps of
$I(\theta,\varphi)$ for a $\sigma$ and $\pi$ transition respectively,
assuming that the QD is surrounded by an infinite medium
made of ZnTe for illustration purpose first. The radial coordinate is equal to
$\sin \theta$ and the polar angle is equal to $\varphi$. The two
radiation patterns differ dramatically, and could be enough to
discriminate light- and heavy-hole emission. Nevertheless, they are
significantly affected by the nanowire geometry and by the
imperfection of the collection objective. 
Note that while the $\pi$ dipole emission does not radiate along the optical axis, it 
does generate light towards larger angles within our experimental numerical aperture, 
which is collected by our objective [see Fig.~\ref{fig:Theory}(d)].
A more precise description
of the radiated field comes from the polarisation of the light
emitted in a given direction. It is projected by the objective onto
a state of polarisation in the $(x,y)$ plane, represented in 
Fig.~\ref{fig:Theory}(c-d) by the green curves which follow the electric
field vector $\bm{E}(\theta,\varphi,t)$ over one optical period. It
results in an ellipse whose aspect ratio changes from a perfect
circle for a pure $\sigma^\pm$ to a single line for a pure linear
polarisation state. Here again, striking differences exist between
$\pi$ and $\sigma$ transitions. One important message of our work is
that this polarisation analysis for a large set of emission
directions provides unambiguous, quantitative informations about the
hole character and the valence band mixing.

\section{Experimental setup and results}

\subsection{Sample fabrication}
\begin{figure}[!ht]
\centering
\includegraphics[width=\linewidth]{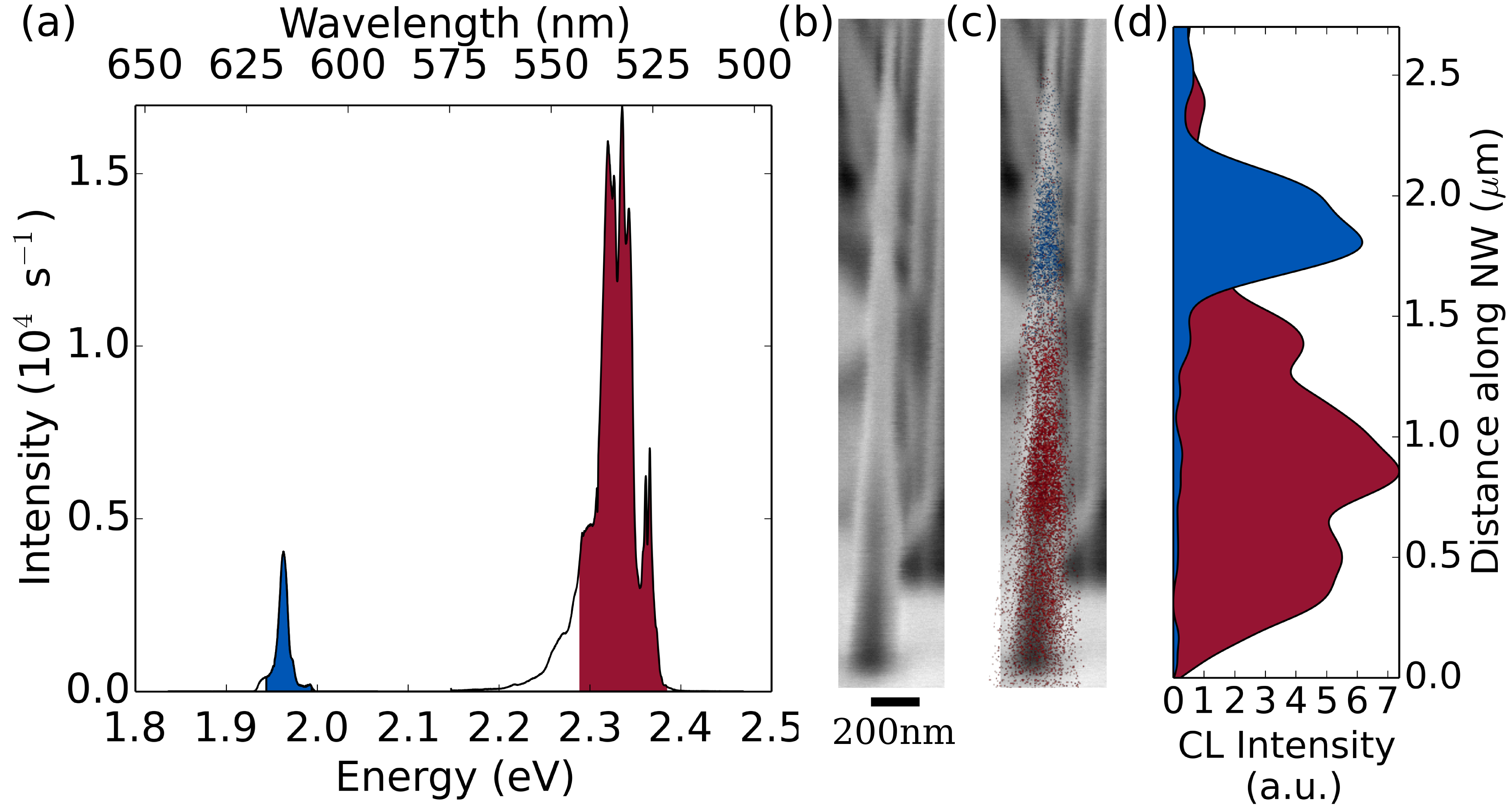}
\caption{
    (a), \textmu PL spectrum of ZnTe core
    and CdTe quantum dot luminescence, along with the integrated spectral range for
    cathodoluminescence imaging. (b), SEM image of the nanowire along with (c),
    spectrally resolved CL image and SEM image superimposed and (d), cut
    profile of the spectrally resolved CL signal along the nanowire axis. The contribution
    of each part of the spectrum is coloured according to the spectrum in
    (a).
    }
\label{fig:Cathodo}
\end{figure}

Our system is a single (Cd,Mn)Te QD inserted along a <111> ZnTe/(Zn,Mg)Te
core/shell nanowire grown by molecular beam
epitaxy (see Supplemental Material \cite{Supplementary}). \cite{Artioli2013, Rueda-Fonseca2014} 
The dot is largely doped with Mn atoms (Mn concentration $\sim$10\%), 
making it a dilute magnetic semiconductor structure.
Due to the nature of the growth, the axial core growth after the quantum dot insertion leads 
to the formation of a ZnTe shell. A final (Zn,Mg)Te shell is further grown on the resulting 
tapered-shape nanowire. 
The wire is standing perpendicular to the
substrate in a region of low nanowire density, allowing its
study by micro-photoluminescence (\textmu PL) without exciting
its neighbours. The PL spectrum  at 5K [Fig.~\ref{fig:Cathodo}(a)] exhibits
an emission peak centered at \unit{2.35}{\electronvolt}, which is
related to the exciton recombination in the ZnTe
core,\cite{Artioli2013, Ferrand2014, Wojnar2014} and a second one at
\unit{1.96}{\electronvolt}, which is attributed to the quantum dot
luminescence. The presence of the Mn atoms significantly broadens the 
emission from the QD because of the magnetization fluctuations randomly 
shifting the exciton line in time through the giant Zeeman effect. 
The nanowire is also studied by low-temperature
cathodoluminescence (CL). The electron beam is along axis $y$. The standard SEM image [Fig.
\ref{fig:Cathodo}(b)] gives access to the geometrical parameters of
the wire. The CL signal [Fig.~\ref{fig:Cathodo}(c-d)] provides
information about the regions from where light is emitted. Most of
the ZnTe luminescence comes from a large region at the base of the
nanowire, while the signal attributed to the quantum dot is well
localized at a height of \unit{1.8}{\micro\meter} from the nanowire
base. We note that the spatial width of this signal is related to
the diffusion of free electron and holes in the nanowire before they
recombine in the dot. It does not correspond to the QD size 
($\sim$\unit{10}{\nano\meter}, measured independently by 
energy-dispersive x-ray spectroscopy.\cite{Rueda-Fonseca2016} 
CL spectroscopy on similar structures confirms that the well-isolated emission line at 
\unit{1.96}{\electronvolt}, which is also well spatially confined, is related to a 
single longitudinal QD, while eventual radial (Cd,Mn)Te structures would emit at a higher 
energy, above \unit{2.1}{\electronvolt}.\cite{Wojnar2016} 
Finally, antibunching experiment on similar emitters without magnetic doping revealed a single 
photon emission with $g^2(0)=0.35$, confirming the 3D confinement of the carriers inside the dot.\cite{StepanovPhD}  
Such experiment could not be performed on our magnetically-doped dots yet because of the line broadening resulting 
from the large magnetic doping of the dot. 

\subsection{Fourier microscopy results}

\begin{figure*}[!htb]
\centering
\includegraphics[width=\linewidth]{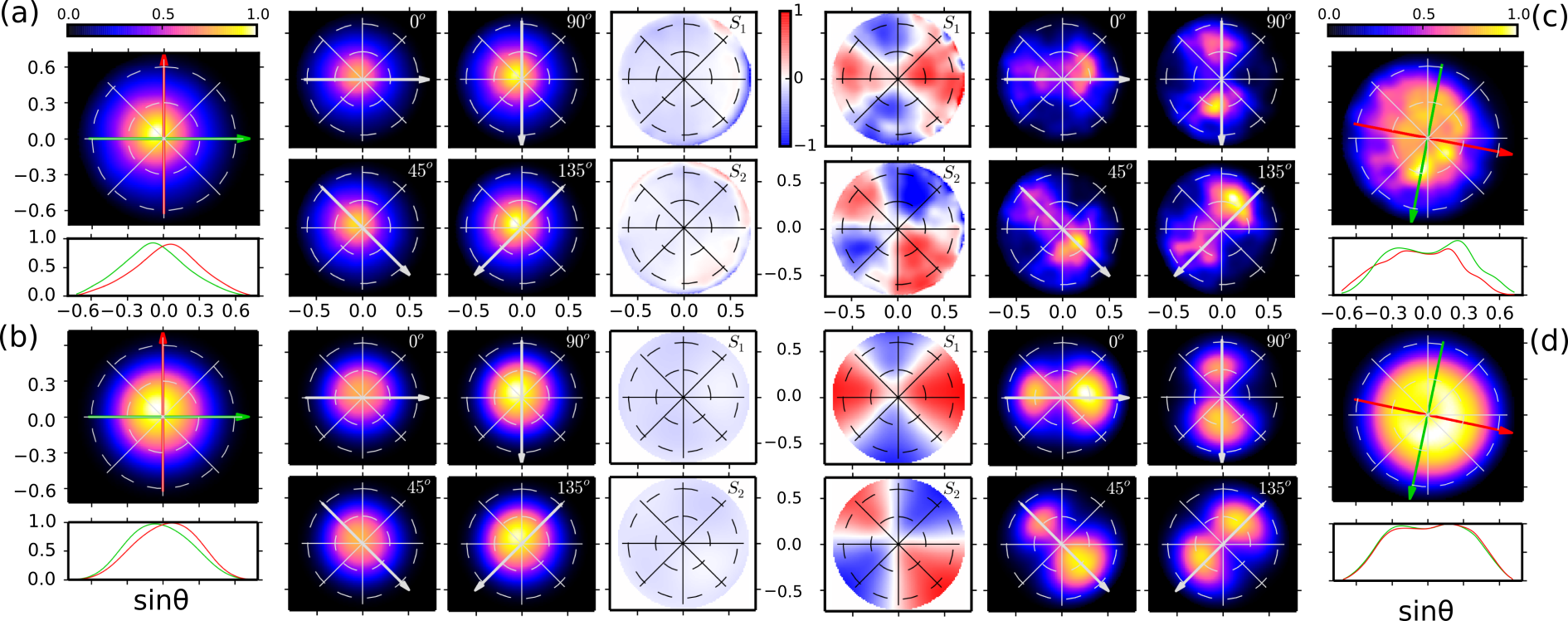}
\caption{
	Coordinate system is the same as in Fig~\ref{fig:Theory}(c). In each panel we
	present the normalized unpolarised far field radiation diagram with two cross
	sections along the direction of the red and green arrows, the linearly polarised
	radiation diagrams at $\{0^o,90^o\}$ and $\{45^o,135^o\}$ sharing the same
	normalization factor (direction of polariser is indicated by gray arrow), and
	the S1 and S2 stokes parameter maps. 
    (a), experimental results for the ZnTe emission line. (b),
    corresponding simulation results ($\rho/\Delta_{LH} = 0.14$, $\sigma/\Delta_{LH}=0.24$,
    $\psi=\unit{112}{\degree}$, $\chi=\unit{186}{\degree}$, or conversely 
    $\rho_0/\Delta_{LH0} = 0.14$, $\hat{\alpha}=\unit{100}{\degree}$, 
    $\hat{\beta}=\unit{5}{\degree}$ and $\hat{\gamma}=\unit{12}{\degree}$).
    (c), experimental results for the QD emission line. 
    (d), corresponding simulations ($\rho/\Delta_{LH} = -0.23$, $\sigma/\Delta_{LH}=-0.06$,
    $\psi=\unit{170}{\degree}$, $\chi=\unit{137}{\degree}$, or conversely 
    $\rho_0/\Delta_{LH0} = -0.23$, $\hat{\alpha}=\unit{20}{\degree}$, 
    $\hat{\beta}=\unit{16}{\degree}$ and $\hat{\gamma}=\unit{150}{\degree}$).   
}
\label{fig:totale}
\end{figure*}

Let us first compare the unpolarised far field radiation pattern of the ZnTe emission
[Fig. \ref{fig:totale}(a)] to the one of the QD [Fig.~\ref{fig:totale}(c)]. Both
present a single lobe of emission whose center is slightly displaced from the
origin. We attribute this off-centering to a geometrical \unit{5}{\degree} tilt of the NW
axis with respect to $z$, different from the previous QD $\beta$ tilt previously 
introduced in the spin Hamiltonian. 
However, we note on the cross-sections that the angular divergence of the ZnTe emission 
is definitely smaller than the QD one, which features a dip at its center. 
This is a first hint for a LH emission from
the QD as the $\pi$ transition reinforces light emission at large angles
[Fig.~\ref{fig:Theory}(d)]. Independent measurements of the objective collection
efficiency show that the latter drops dramatically for $\sin
\theta \geq 0.45$ (see Supplemental Material (\cite{Supplementary}). 
This does not affect the comparison between the two lines,
but prevent us to attempt a direct comparison with the calculated patterns of
Fig.~\ref{fig:Theory}(c-d). Another issue is that the ZnTe emission takes place in
a region where the nanowire diameter is such that it strongly guides light along
the axis. Even more, due to the nanowire cone shape, the guided mode waist
increases adiabatically, and hence its angular divergence decreases - a
mechanism which is exploited in photonic wires to maximize light collection from
single QD.\cite{claudon2010highly,Bulgarini2014} On the contrary, at the QD
location the nanowire diameter is too small to allow an efficient guiding
effect.

The difference between ZnTe and QD emission is dramatically highlighted when
comparing their linearly polarised radiation patterns. In the case of ZnTe emission 
they remains similar to the unpolarised one,
whatever the polariser direction. On the contrary, the QD emission patterns break the
revolution symmetry around $z$. We observe two lobes, off-centered and
symmetrically placed on both sides of the optical axis along the direction of
polarisation. The lobe intensity decreases at large $\theta$ angle because of
the loss of collection efficiency otherwise it would be maximum at the edge of
the image as in Fig.~\ref{fig:Theory}(b). We also note that there is a $\sim$20\%
intensity imbalance between the two lobes. 
A convenient way to reinforce the information about the polarisation state of the
far field, without the influence of its intensity, is to plot the Stokes
parameters \cite{McMasterMcMaster_MatrixRepresentationPolarization_61} $S_1 =
(I_{0}-I_{90})/(I_{0}+I_{90})$ and  $S_2 = (I_{45}-I_{135})/(I_{45}+I_{135})$,
where $I_\alpha$ is the far field intensity for a linear polariser set at angle
$\alpha$. The degree of linear polarisation (DLP) is equal to $S_1^2+S_2^2$. In
the case of the ZnTe line, both Stokes parameters are homogeneous and very close
to 0, with a DLP averaged over all measured directions of light $\sim$2\%. On
the contrary, the QD Stokes parameters are varying with $\varphi$ from very large
positive values to very negative ones, displaying a characteristic 4-quadrant
symmetry. S2 is similar to S1 rotated by \unit{45}{\degree}. The average DLP is
$\sim$40\%. This is a direct consequence of the polarisation properties
sketched in green in Fig.~\ref{fig:Theory}(c-d).

Experimental results were compared to far field patterns derived from
simulations of the electromagnetic field in the whole nanowire
structure by a finite element software and taking into account all the 
experimental imperfections of our imaging setup 
(see Supplemental Material \cite{Supplementary}). 
The emitter is modeled by an oscillating dipole $\bm{d}=d_x\bm{x}+d_y\bm{y}+d_z\bm{z}$. 
The dipole matrix elements $d_i$ are determined by diagonalizing the
hole Hamiltonian including valence band mixing, and considering all
possible transitions between the ground hole state and the electron
states. For the ZnTe core [Fig.~\ref{fig:totale}(b)], we sum
the intensities coming from dipoles emitting at different positions
along the nanowire axis, with weights corresponding to the CL
intensity in Fig.~\ref{fig:Cathodo}(d). 
For the QD [Fig.~\ref{fig:totale}(d)], the dipole is fixed at
the QD position. The agreement with
experimental data is very good. Looking at the unpolarised QD pattern,
the effects of the valence band mixing rates $\sigma/\Delta_{LH}$ and
$\rho/\Delta_{LH}$ are entangled.
However, the unbalanced intensity lobes (central pannel) reflect
$\sigma/\Delta_{LH}$ with its phase, while
the linear polarisation observed on S1 close to the optical axis
reflects $\rho/\Delta_{LH}$ and its phase. The mixing rates are
small, hence the QD hole state is essentially ($\sim$97\%)
LH. Yet non zero valence band mixing is necessary to
properly explain the fine features found in the experimental data.

\subsection{Magneto-optical spectroscopy}

\begin{figure}[!ht]
\centering
\includegraphics[width=0.9\linewidth]{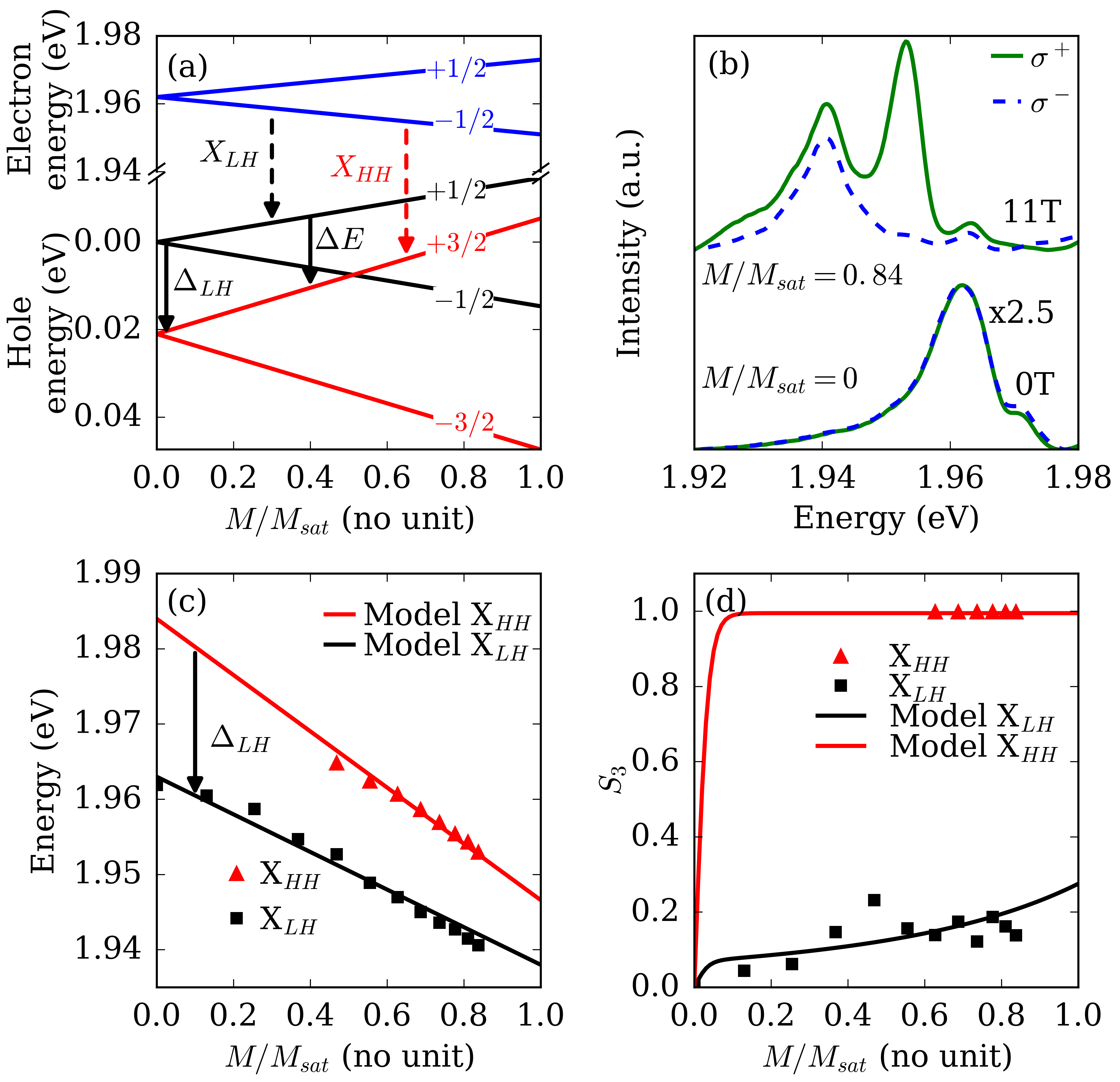}
\caption{
	(a),
Zeeman diagram of the electron and hole confined levels split by the exchange
interactions with Mn atoms. The dashed arrows labeled $X_{LH}$ (resp. $X_{HH}$) corresponds to the LH 
(resp. HH) exciton transitions. 
The low-energy transition involving a LH creates a $\ket{-1/2}$ electron in the 
conduction band and a $\ket{+1/2}$ hole in the valence band, hence it is linearly polarized 
along the $z$-axis. The low-energy transition involving a HH creates a $\ket{-1/2}$ 
electron in the conduction band and a $\ket{+3/2}$ hole in the valence band, 
hence it is $\sigma^{+}$-polarized. 
(b), Photoluminescence spectra of the QD emission line recorded at 10K in
zero field and at B=11T (corresponding to a QD magnetic moment
$M/M_{sat}=0.84$). The magnetic field is applied along the nanowire
axis oriented parallel to the optical axis. Both circular
polarisations ($\sigma^+$ in green and $\sigma^-$ in dashed blue) are shown.
The intensity of the zero field spectrum has been multiplied by 2.5. 
(c), Zeeman shift of the LH and HH exciton
lines proportional to the QD magnetic moment $M/M_{sat}$. The
continuous lines correspond to the theoretical shift expected for a
(Cd,Mn)Te quantum dot with a Mn concentration of 10\% (see text).
(d), Circular polarisation rates of the LH and HH exciton lines under magnetic field. The continuous lines
corresponds to the expected values taking into account the valence band mixing induced by a constant perturbation term
$|\sigma|=3.9$ meV.
} \label{fig:zeeman}
\end{figure}

The analysis of the QD lines under magnetic field (Zeeman
splitting and circular polarisation degree) provides another way to
discriminate between LH and HH excitons. In our experiment, the magnetic field 
is applied along the NW axis (Faraday configuration). 
In non magnetic quantum dots with moderate exciton Landé factors,
large magnetic fields are usually required in order to lift
exciton spin manifold degeneracy. In (Cd,Mn)Te magnetic quantum dots, Mn atoms
introduce localized spins $S=5/2$ randomly distributed in the dot.
The large exchange interaction between Mn spins and confined
carriers (electrons and holes) induces the so-called giant Zeeman
splitting of the exciton lines at low temperature (effective
Landé factor $\sim$100), which scales proportionally to the normalized quantum dot
magnetic moment $M/M_{sat}$ 
(see section I in the Supplemental Material \cite{Supplementary}).\cite{Clement2010} 
Even higher values are obtained with colloidal quantum dots.\cite{Beaulac2009} 
In this case the strong anisotropy due to the wurtzite crystal structure enforces 
a heavy-hole character. 
Fig.~\ref{fig:zeeman}(a) shows the Zeeman energy diagram of electrons and
holes in an (Cd,Mn)Te/ZnTe QD with a 10\% Mn
concentration under a magnetic field applied
along the $z$ axis and assuming $\Delta_{LH}<0$. At low
temperature and for magnetic fields larger than a fraction of \tesla,
photo-carriers relax to the lowest energy levels ($\ket{-1/2}$ for
electrons and $\ket{+1/2}$ for holes). This induces both a red shift
of the LH exciton line (black dashed arrow in Fig.~\ref{fig:zeeman}(a),
and a strong linear polarisation along $z$ ($\pi$ polarisation),
according to the optical selection rules recalled in Fig.~\ref{fig:Theory}(a).
With the opposite assumption of a HH ground state,
we would expect a $\sigma^+$ polarisation.

In our experimental configuration, the magnetic field is applied along 
the NW axis which still stands along the optical axis (Faraday configuration), and 
the QD luminescence is collected over an angular range of NA = 0.4. 
Figure \ref{fig:zeeman}(b) shows the quantum dot emission spectra resolved in
$\sigma^+$ and $\sigma^-$ polarisations for an applied magnetic field of
\unit{0}{\tesla} and \unit{11}{\tesla}. The emission at zero field is dominated
by the line at \unit{1.962}{\electronvolt} previously studied in far-field. At
\unit{11}{\tesla} (corresponding to $M/M_{sat}\simeq 0.84$ at
$T=$\unit{10}{\kelvin}), the spectrum consists in two main lines at
\unit{1.94}{\electronvolt} and \unit{1.953}{\electronvolt}. 
Both spectra present a weak satellite line at higher energy whose origin is unclear. 
The two lines are probably not related and a rather small Zeeman shift is observed 
(see Fig.~\ref{fig:magnetooptics} in Appendix \ref{app:magnetooptics}). 
It could be related to the luminescence from the substrate, or from parasitic growth 
which takes place between the nanowires.
The Zeeman shift of the two main lines as a function of the QD magnetic moment 
$M/M_{sat}$ is reported in Fig.~\ref{fig:zeeman}(c). 
The first line, observed for all magnetic field values,
presents a redshift proportional to $M/M_{sat}$. The shift at saturation is
\unit{25}{\milli\electronvolt} as expected for a LH exciton in a
Cd$_{0.9}$Mn$_{0.1}$Te magnetic QD (see Appendix \ref{app:magnetooptics}). 
It shows very moderate circular polarisation imbalance, 
as expected for a LH emission which is ideally $\pi$ polarised. 
It confirms the LH exciton character of this line, as claimed 
by the analysis of farfield patterns in zero magnetic field. The second line is
clearly present for fields larger than \unit{6}{\tesla} ($M/M_{sat}>0.6$). It is
strongly $\sigma^+$ polarised. It also redshifts proportionally to $M/M_{sat}$
with a shift at saturation of \unit{37}{\milli\electronvolt}. This value is
definitely larger than the maximum possible LH Zeeman shift, whatever the Mn
content, of \unit{25}{\milli\electronvolt}. For these two reasons, we ascribe
this line to the $\ket{-1/2}\rightarrow\ket{3/2}$ transition [red arrow in Fig.~\ref{fig:zeeman}(a)] 
associated to the HH exciton. 
The saturation shift is lower than the value expected for a
heavy-hole fully confined inside the dot (\unit{51}{\milli\electronvolt}),
suggesting a rather weak confinement of the HH excited state in the dot.  At 
low field the behaviour of the line is complicated due to coupling to LH states. 
By extrapolating the Zeeman shift of the transition observed at high field down to 
$B=$\unit{0}{\tesla} we obtain a splitting $\Delta_{LH}=-\unit{22}{\milli\electronvolt}$.
We want to stress here that the difference in the polarization of the two lines is a strong 
indication of the different nature of the hole involved in each transition, and rules out 
the possible emission from regions with different Mn content. 

Information about the valence band mixing can be retrieved by a detailed
analysis of the third Stokes parameter
$S_3=\frac{I_{\sigma^+}-I_{\sigma^-}}{I_{\sigma^+}+I_{\sigma^-}}$, plotted in
Fig.~\ref{fig:zeeman}(d) $S_3$. The
strong $\sigma^+$ polarisation of the HH exciton line fully corresponds
to the radiative recombination of a $\ket{-1/2}$ electron with a $\ket{+3/2}$
hole, and the selection rules given in
Fig.~\ref{fig:Theory}(a). Surprisingly, one can note that the LH exciton
line is also partially circularly polarised at large magnetic field (for
$M/M_{sat}>0.5$). This partial circular polarisation results from the hole 
mixing induced by the perturbation term $\sigma e^{-i\chi}$ 
(see Supplemental Material \cite{Supplementary}). 
The black line in Fig.~\ref{fig:zeeman}(d) give the theoretical 
variation of $S_3$ for the LH transition assuming a value of 
$\sigma=$\unit{3.9}{\milli\electronvolt} independent of the magnetic field and 
taking into account the ratio in  collection efficiency of $\pi$ and $\sigma$  
polarised emission $f_{\sigma\pi}=1.8$  into the objective lens. The
thermalization of the holes between the Zeeman levels has been added in order to
get vanishing circular polarisation in zero field. Using the hole energy levels 
deduced from the Zeeman shifts of the exciton lines it yields 
$\left(\frac{\sigma}{\Delta_{LH}}\right)^2 \simeq$3\%, in agreement with the 
the simulations of the zero field emission diagram. Due to the large value of
$f_{\sigma \pi}$, the mixing terms do not affect $S_3$
for the HH exciton at large magnetic field: the red line in
Fig.~\ref{fig:zeeman}(d) displays the circular polarisation expected for the heavy
hole exciton.

\section{Discussion}


The polarized emission diagram of the ZnTe nanowire unambiguously reveals the HH
character of the exciton, with a small mixing. This is expected
\cite{Ferrand2014} from the presence of the (Zn,Mg)Te shell with a smaller lattice
parameter; the expected redshift was experimentally confirmed,\cite{Wojnar2012,Artioli2013} 
and the heavy hole character was deduced from the giant Zeeman
effect and the circular polarization observed with a (Zn,Mn)Te core. Using the
composition profile obtained from Ref.~\citenum{Rueda-Fonseca2016} for a nanowire
from the same sample, and analytical expressions of the strain-induced splitting,
\cite{Ferrand2014, Artioli2013} we expect
$\Delta_{LH}\simeq+$\unit{30}{\milli\electronvolt} and a ZnTe emission energy
around \unit{2350}{\milli\electronvolt} in agreement with
Fig.~\ref{fig:Cathodo}(a). This is consistent with the
\unit{50}{\milli\electronvolt} value deduced from  the anisotropy of the giant
Zeeman effect in (Zn,Mn,Te)/(Zn,Mg,Te) nanowires including a larger Mg
content.\cite{Szymura2015} The mismatch between CdTe and ZnTe is opposite, so
that we expect a LH character for a hole confined in a core-shell nanowire
\cite{Ferrand2014} or an elongated dot.\cite{Zielinski2013} For InAs/InP, a
local splitting $\Delta_{LH}\simeq$\unit{-100}{\milli\electronvolt} is
calculated \cite{Zielinski2013} at the center of a cylinder-shape QD with an 
aspect ratio 2, but due to an inhomogeneous strain the HH state is
$\sim$\unit{25}{\milli\electronvolt} above the ground state. In the present
case, the aspect ratio is of the same order, the mismatch slightly larger, but
the CdTe-ZnTe valence band offset is small so that another competition is
expected with heavy-holes confined in the shell due to the shear strain around
the dot.\cite{Ferrand2014} A full calculation of $\Delta_{LH}$ is beyond the scope
of this paper, but our measured value is of the right order or magnitude.

We measure a value of the mixing term $\rho_0\simeq 5$~meV. Larger
values have been reported for self-assembled (Stranski-Krastanov)
quantum dots.\cite{Tonin2012} Actually, an important feature is the
symmetry of the principal axis, and switching from the $<001>$ to
the $<111>$ orientation dramatically reduces in-plane asymmetry
expected,\cite{Singh2009} and measured.\cite{Kuroda2013, Juska2013} Our dots
are embedded in nanowires grown along the $<111>$ axis. However the
section of such nanowires easily feature some ellipticity, of the
order of a few $\%$, easily detected on the shell \cite{Rueda-Fonseca2016}
(although well beyond the resolution on the dot). A first evaluation
of the strain in an elongated (aspect ratio $\sim~2$) ellipsoid with
some in-plane ellipticity can be done using the Eshelby calculation:
\cite{Eshelby1957,Eshelby1959} using the Bir-Pikus Hamiltonian, an ellipticity of
$5\%$ gives the right order of magnitude for $\rho_0$. Note that the
same orientation of the ellipticity will change the phase of $\rho$
with the sign of the mismatch, as observed between the dot and the
ZnTe core. Finally, writing the stiffness matrix with the $<111>$ axis as
$z$-axis \cite{Ferrand2014} reveals that the presence of an in-plane
shear ($\varepsilon_{xx}-\varepsilon_{yy}$) induces an axial shear
stress, and hence a strain $\varepsilon_{yz}$, of similar order of
magnitude.  This strain gives rise to a non-zero $\sigma$, which is
not a tilt, but can be compensated by a tilt. Such a term is
evidenced on the emission diagram of the quantum dot, while it is
partly screened by guiding effect in the case of the ZnTe emission.

We have performed measurements on a set of $\sim40$ NWQDs from the same sample
mechanically deposited on a substrate. As they lie horizontally and close to a
reflecting substrate which significantly disturbs the far-field radiation
pattern, we could not perform the same characterization as for the as-grown
wire. Nevertheless polarization studies reveal that the sample is actually close
to the threshold between light-hole and heavy-hole ground
state,\cite{Zielinski2013} thus leading to a large dispersion in emission
properties. A majority of the QDs ($\sim 70\%$) emit light linearly polarized
along the NW axis, which in our system is a good indication of a LH ground
state. The remaining $30\%$ emit light polarized perpendicularly to the NW axis,
thus clearly indicating a HH character.

To conclude, complementary experiments on a single nanowire unambiguously 
demonstrate a LH exciton emission. They allow to evaluate the splitting 
$\Delta_{LH}$ as  well as the valence band mixing parameters. The results are in 
agreement with the predictions of strain effects due to the presence of a shell 
around the wire. 
We demonstrate that valence band engineering through strain and 
confinement is possible using bottom up approach and semiconducting nanowire growth.

\begin{acknowledgments}
	We acknowledge the help of Institut N\'eel optical engineering team (CL, Fabrice Donatini).
	This work was supported by the French National Research Agency under contracts ANR-11-BS10-013, 
	ANR-15-CE24-0029 and ANR-10-LABX-51-01, and  Institut Universitaire de France.
\end{acknowledgments}

\appendix

\section{Magneto-optical spectroscopy}
\label{app:magnetooptics}
\begin{figure}[!ht]
    \begin{center}
        \includegraphics[width=\linewidth]{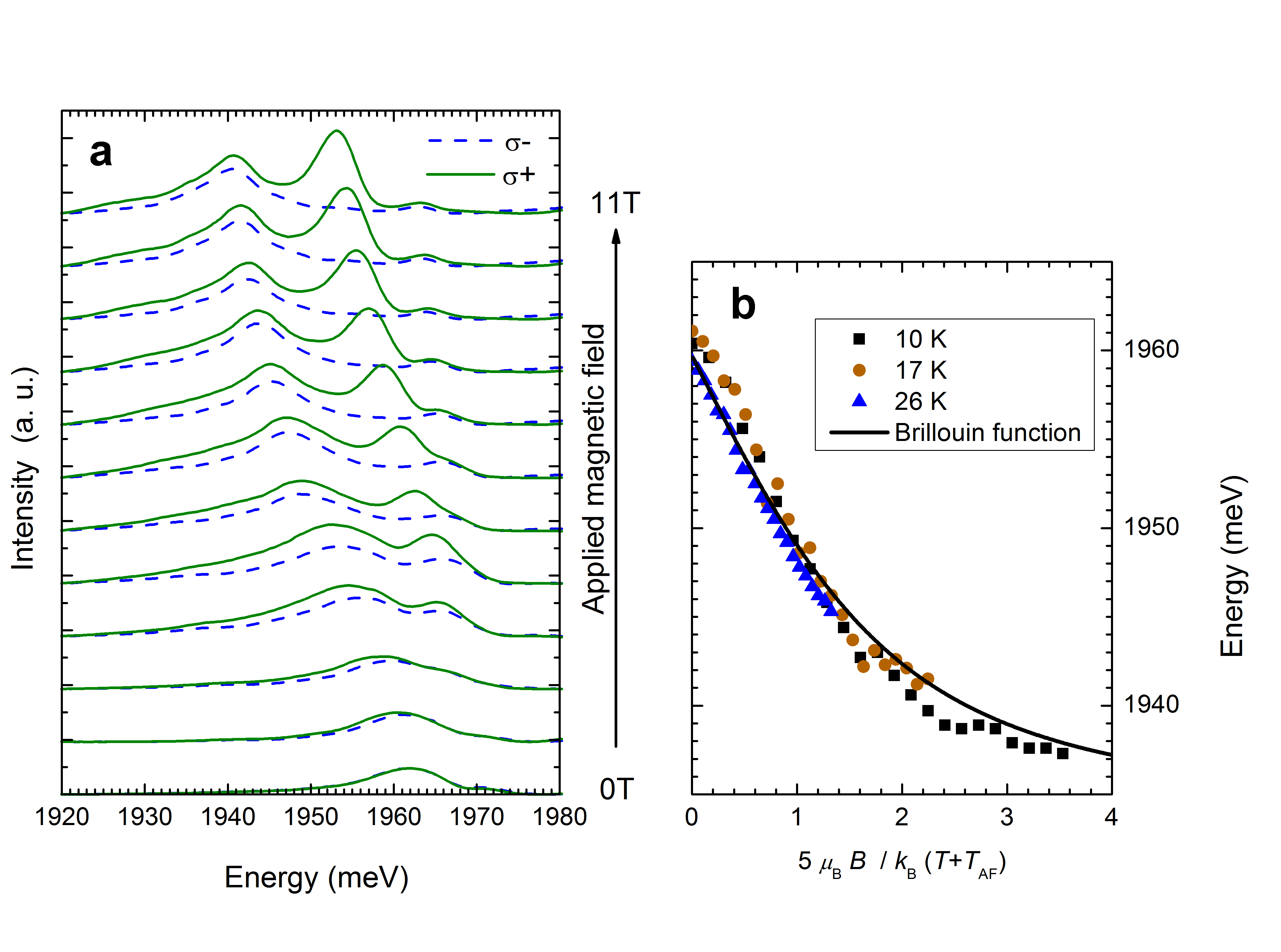}
    \end{center}
    \caption{\textbf{Magneto-optical spectroscopy}.\\
        (a) Photoluminescence spectra at $T=10~$K with a magnetic field applied along the nanowire axis, 
        from $B=0$ to 11~T by steps of 1~T. The spectra at 0 and 11~T are displayed in Fig.~4a of main text. 
        (b) Plot of the position of the low energy line as a function of $5 \mu_B B / k_B (T+T_{AF}$)}
    \label{fig:magnetooptics}
\end{figure}

Figure \ref{fig:magnetooptics}a displays photoluminescence spectra
recorded at different values of the magnetic field applied along the
nanowire axis. Only the extreme field values are plotted in Fig. 4a
of the main text. Most salient features are the intense low energy
line, which is present at zero field and displays a continuous
red shift when increasing the intensity of the field. Fig.
\ref{fig:magnetooptics}b, the position of this line is plotted for
different values of the applied magnetic field and three values of
the temperature (spectra of Fig.~\ref{fig:magnetooptics}a, and spectra
at two other temperatures, not shown), as a function of $5 \mu_B B /
k_B (T+T_{AF})$. The phenomenological parameter $T_{AF}$ describes
the antiferromagnetic interactions in Cd$_{0.9}$Mn$_{0.1}$Te, see
above. The coincidence of the three sets of data confirms that the
shift is due to the giant Zeeman effect, and follows a Brillouin
function with a shift at saturation equal to \unit{25}{\milli\electronvolt} (solid line).
Note that this line exhibits only a small circular polarization. The
second salient feature is the strongly $\sigma^+$-polarized line,
visible at high field only. A weaker line is observed at high field,
with a small Zeeman shift, which we did not identify (the nanowire
is still on the substrate and parasitic growth takes place between
the nanowires). Finally, the spectra at low fields appear as quite
complex, but this is expected by the proximity of several hole
sublevels at these field values (in particular, $\ket{+3/2}$ and $\ket{-1/2}$ which
are expected to (anti)-cross at these field values, see Fig.~4b of
the main text, and probably excited states of the $\ket{+1/2}$ hole).

\section{Theoretical background}\label{app:theo}

\subsection{Anisotropy of the holes}

In a bulk semiconductor with the zinc-blende or diamond structure,
the hole quadruplet is degenerate (representation $\Gamma_8$ of the
cubic group). The presence of strain or confinement lifts this
degeneracy. If a principal axis of symmetry exists, the hole
Hamiltonian is usually written, in the $\{\ket{3/2};\ket{1/2};
\ket{-1/2};\ket{-3/2}\}$ basis quantized along this axis, as:
\begin{equation} \label{VBMmatrix}
\cal{H} =
\begin{pmatrix}
-\frac{1}{2}\Delta_{LH} & -\sigma e^{-i\chi} & \rho e^{-2i\psi} & 0 \\
-\sigma e^{i\chi} & \frac{1}{2}\Delta_{LH} & 0 & \rho e^{-2i\psi} \\
\rho e^{2i\psi} & 0 & \frac{1}{2}\Delta_{LH} & \sigma e^{-i\chi} \\
0 & \rho e^{2i\psi} & \sigma e^{i\chi} &-\frac{1}{2}\Delta_{LH}
\end{pmatrix}
\end{equation}

If $\Delta_{LH}$ is larger than $\rho$ and $\sigma$, the two Kramers
doublet defined by this Hamiltonian are usually called light-hole
and heavy-hole (even if these terms are not always justified), with
$\Delta_{LH}$ the light-hole to heavy-hole energy splitting (the
exact value of the splitting is
$\sqrt{\Delta_{LH}^2+4\rho^2+4\sigma^2}$).

Two well known examples of this Hamiltonian are the Luttinger-Kohn
Hamiltonian which describes the hole states in the vicinity of the
valence band maximum, and the Bir-Pikus Hamiltonian which describes
the coupling to a uniform strain.\cite{Fishman}

We want to stress here that this Hamiltonian $\cal{H}$ is much more
general: if the hole quadruplet is isolated, $\cal{H}$ can be
considered as a spin Hamiltonian, \emph{i.e.}, as an effective
Hamiltonian operating within the quadruplet.\cite{AbragamBleaney} 
Some care must be taken when excited states have to be considered.\cite{Luo2015} 

Moreover, this effective Hamiltonian can be built as a linear
combination of the successive powers of a pseudo-spin, with real
coefficients. In the case of a $J=3/2$ quadruplet, powers of the $J$
operators up to 3 are enough, and in the absence of an applied
magnetic field, only even powers have to be considered in order to
fulfill the Kramers degeneracy. As a result, the most general spin
Hamiltonian which describes the quadruplet at the top of the valence
band can be written $\bm{J}.\bm{A}.\bm{J}$, where the
vectorial operator $\bm{J}$ is the (pseudo)-moment and
$\bm{A}$ is a real $3\times3$ matrix. In addition, due to the
commutation rules of $\bm{J}$, $\bm{A}$ is symmetric. Note
that the contribution of order zero to the spin Hamiltonian is
redundant with the trace of $\bm{A}$; both represent a rigid
shift of the quadruplet, and a proper choice of the zero of energy
allows us to set the trace of $\cal{H}$ (hence that of $\bm{A}$)
to zero, as done in Eq.~\ref{VBMmatrix}.

Using the $4\times4$ matrices representing the second powers of
$\bm{J}$ in the $|\frac{3}{2}\rangle$, $|\frac{1}{2}\rangle$,
$|-\frac{1}{2}\rangle$ and $|- \frac{3}{2}\rangle$ basis of the
$\Gamma_8$ quadruplet, with the third axis $z$ as the quantization
axis\cite{Fishman}, the matrix elements of
$\cal{H}$ are
\begin{eqnarray} \label{AtoH}
Tr(\cal{H}) &=& \frac{5}{4}(A_{xx}+A_{yy}+A_{zz})=0\nonumber  \\
\frac{1}{2}\Delta_{LH} &=& \frac{1}{2}A_{xx}+\frac{1}{2}A_{yy}-A_{zz}\nonumber  \\
\rho e^{-2i\psi} &=& \frac{\sqrt{3}}{2}(A_{xx}-A_{yy}-2iA_{xy})\nonumber\\
\sigma e^{-i\chi} &=& -\sqrt{3}(A_{xz}-iA_{yz})
\end{eqnarray}

Expressions of the matrix elements of Eq.~\ref{VBMmatrix} (hence
those of $\bm{A}$) for the Luttinger-Kohn of Bir-Pikus Hamiltonian
are generally expressed in the cubic basis.\cite{BirPikus} However
other axes can be chosen; for instance, in the case of a nanowire
oriented along the $<111>$ axis, as in the present study, it is
useful to choose this axis as the $z$ quantization axis.\cite{Ferrand2014} 
If $\cal{H}$ represents the coupling to a
uniform strain (the Bir-Pikus Hamiltonian) for an isotropic system,
$\Delta_{LH}$ is proportional to the axial shear strain
$(\frac{1}{2}\varepsilon_{xx}+\frac{1}{2}\varepsilon_{yy}-\varepsilon_{zz})$,
$\rho e^{-2i\psi}$ to the shear strain in the $xy$ plane,
$\frac{\sqrt{3}}{2}(\varepsilon_{xx}-\varepsilon_{yy}-2i\varepsilon_{xy})$,
and $\sigma e^{-i\chi}$ to the combination
$-\sqrt{3}(\varepsilon_{xz}-i\varepsilon_{yz})$ of the shear strains
in planes containing $z$. However, it must be kept in mind that such
a spin Hamiltonian is general and can describe other features
governing the hole states, such as for instance the inhomogeneous
strain expected in a quantum dot, or the effect of a confinement
potential with a low symmetry.

Now, as the matrix $\bm{A}$ is real and symmetric, a mere
rotation makes it diagonal, with the three (real) eigenvalues on the
diagonal. Using these eigenaxes ($x_0$, $y_0$, $z_0$), the spin
Hamiltonian still writes $\bm{J}.\bm{A}.\bm{J}$, and it
still develops as in Eq.~\ref{VBMmatrix},  but Eq.~\ref{AtoH} shows
that now all matrix elements are real (including $\rho$), and
$\sigma=0$:

\begin{equation} \label{VBMmatrix0}
\cal{H} =
\begin{pmatrix}
-\frac{1}{2}\Delta_{LH0} & 0 & \rho_0& 0 \\
0 & \frac{1}{2}\Delta_{LH0} & 0 & \rho_0 \\
\rho_0 & 0 & \frac{1}{2}\Delta_{LH0} & 0 \\
0 & \rho_0 & 0 &-\frac{1}{2}\Delta_{LH0}
\end{pmatrix}
\end{equation}

with $A_{z_0z_0}=\frac{1}{3}\Delta_{LH0}$,
$A_{x_0x_0}=-\frac{1}{6}\Delta_{LH0}+\frac{1}{\sqrt{3}}\rho_0$ and
$A_{y_0y_0}=-\frac{1}{6}\Delta_{LH0}-\frac{1}{\sqrt{3}}\rho_0$. If
the $z_0$ axis is chosen so that $A_{z_0z_0}$ is the eigenvalue with
the largest absolute value, then $\rho_0<|\Delta_{LH0}|/2\sqrt{3}$;
the two Kramers doublet will be considered as light holes and heavy
holes quantized along $z_0$, with some mixing due to $\rho_0$.

To sum up, the most general spin Hamiltonian describing the hole
states is given by Eq.~\ref{VBMmatrix}, where diagonal elements are
real numbers but non-diagonal elements are complex numbers.
Diagonalizing the corresponding matrix $\bm{A}$ determines a
rotation (\emph{i.e.}, three Euler angles) to a new set of axes
where $\Delta_{LH0}$ but also $\rho_0$ are real numbers and
$\sigma$=0, Eq.~\ref{VBMmatrix0}.

\begin{figure}[!htb]
	\begin{center}
		\includegraphics[width=\linewidth]{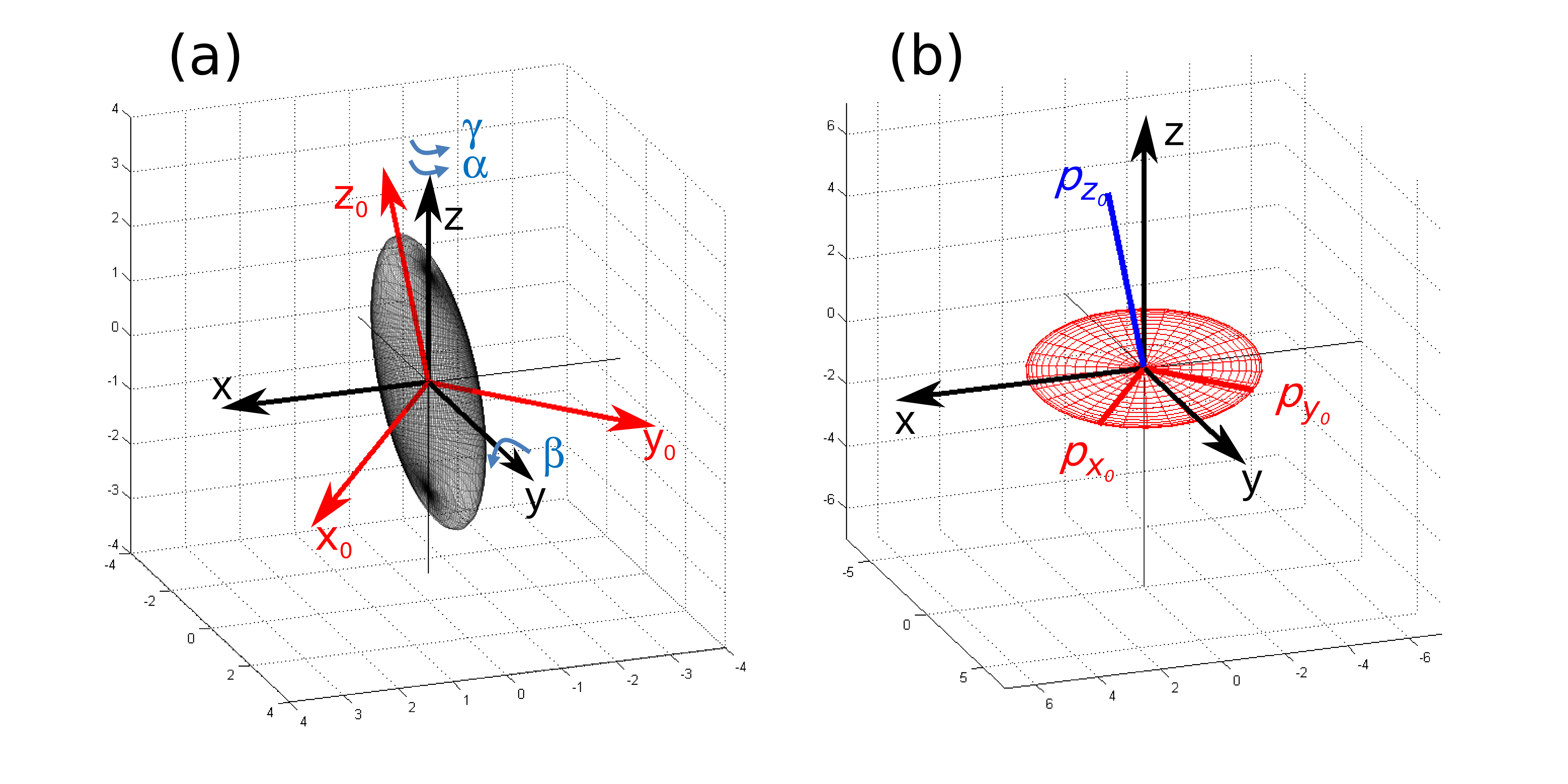}
	\end{center}
	\caption{\textbf{Light Hole exciton dipole radiation}  (a)
		Principal axes $(x_0,y_0,z_0)$ of the ellipsoid defined by
		Eq.~\ref{ellipsoid} and calculated for $\psi=40\degree$,
		$\chi=110\degree$, $\rho/\Delta_{LH}=0.1$, $\sigma/\Delta_{LH}=0.3$
		and $E_0/\Delta_{LH}=2$. They are obtained by 3 rotations with the
		Euler angles $\hat{\alpha}=40 \degree$, $\hat{\beta}=19\degree$ and
		$\hat{\gamma}=0\degree$. (b) Orientation of the 3
		elementary orthogonal linear dipoles associated to the light hole
		exciton. They are oriented along the axes $(x_0,y_0,z_0)$. The $\pi$
		(resp. $\sigma$) polarized exciton transitions are associated to the
		dipole along $z_0$ (resp. $x_0$ and $y_0$). The figure is calculated
		for the same Euler angles as above and for the hole anisotropy
		parameters $\epsilon=0.214$ ($\rho_0/\Delta_{LH0}=0.0406$). The
		lengthes of the colored lines are proportional to the dipole
		magnitudes $d_{x_0}$, $d_{y_0}$ and $d_{z_0}$ (see text).}
	\label{fig:dip}
\end{figure}

As the principal axes of $\bm{A}$ are those where the
Hamiltonian $\cal{H}$ is real with a vanishing $\sigma$, so that the
hole states are most simple to write, it is useful to write the
effect of a rotation on $\bm{A}$  and $\cal{H}$. We apply to
$\bm{A}$ or $\cal{H}$ the rotation $\exp (-i \hat{\gamma}
J_z)~\exp (-i \hat{\beta} J_y)~\exp (-i \hat{\alpha} J_z)~$ defined
by three Euler angles $(\hat{\alpha}, \hat{\beta}, \hat{\gamma})$,
see Fig.~\ref{fig:dip}. It is particularly interesting - and
technically simpler - to consider small values of $\hat{\beta}$, so
that we keep only the first order in $\hat{\beta}$ or $\sin
\hat{\beta}$. In this case, $(\hat{\alpha}+\hat{\gamma})$ is the
total angle of rotation around the $z$ axis (which has an effect
only if the system does not feature circular symmetry, with non
vanishing $\rho_0$ and different values of $A_{x_0x_0}$ and
$A_{y_0y_0}$), while $\hat{\beta}$ is the angle of the tilt and
$\hat{\gamma}$ its direction. The rotation matrix in real space is

\begin{equation} \label{Euler}
\cal{R} =
\begin{pmatrix}
\cos (\hat{\alpha}+\hat{\gamma})& -\sin (\hat{\alpha}+\hat{\gamma}) & \sin \hat{\beta} \cos \hat{\gamma}\\
\sin (\hat{\alpha}+\hat{\gamma}) & \cos (\hat{\alpha}+\hat{\gamma}) & \sin \hat{\beta} \sin \hat{\gamma}\\
-\sin \hat{\beta} \cos \hat{\alpha} & \sin \hat{\beta} \sin
\hat{\alpha} & 1
\end{pmatrix}
\end{equation}

and the matrix elements of $\bm{A}$ in the laboratory frame are
obtained by a straightforward calculation of $\cal{R} \bm{A}$
${}^t \cal{R}$. $\cal{H}$ is derived using Eq.~\ref{AtoH}, or
calculated directly using the rotation $\exp (-i \hat{\gamma}
J_z)~(1-i \hat{\beta} J_y)~\exp (-i \hat{\alpha} J_z)~$, where the
operator of rotation around $y$ (the tilt) is linearized. The result
is quite simple. The hamiltonian assumes the form of
Eq.~\ref{VBMmatrix}, where:

\begin{eqnarray} \label{tilt}
\frac{1}{2}\Delta_{LH} &=& \frac{1}{2}\Delta_{LH0}\nonumber  \\
\rho e^{-2i\psi} &=& \rho_0 e^{-2i(\hat{\alpha}+\hat{\gamma})}\nonumber\\
\sigma e^{-i\chi}&=& \hat{\beta} (\frac{\sqrt{3}}{2}
\Delta_{LH0}+\rho_0 e^{-2i\hat{\alpha}}) e^{-i\hat{\gamma}}
\end{eqnarray}

These expressions have been used to fit the experimental data.

\subsection{Geometrical visualizing of the anisotropy tensor}

Since in the case of the Bir-Pikus Hamiltonian, the anisotropy
matrix $\bm{A}$ is proportional to the strain tensor (possibly
weighted by the deformation potential parameters), it is often quite
illuminating to pursue the analogy and to consider an ellipsoid
defined in real space using the matrix $\bm{A}$, as an easy way
to view the anisotropy within the hole quadruplet. One possibility
is to consider a sphere of radius unity, to which we apply a strain
$-A_{ij}/E_0$, with an arbitrary scaling factor $E_0$. If $E_0$ is
larger than the largest eigenvalue of $\bm{A}$, the result is an
ellipsoid which admits the $x_0$, $y_0$, $z_0$ eigen-axes of
$\bm{A}$ as principal axes, with a half-axis length
$1-A_{x_0x_0}/E_0$ along $x_0$, and so on.

To first order in $A_{ij}/E_0$, the equation of such a solid is
\begin{equation} \label{ellipsoid}
\sum_{i,j} x_i ( \delta_{ij} + 2A_{ij}/E_0)x_j=1
\end{equation}
where the $x_i$'s are the coordinates $x$, $y$, $z$ of an arbitrary
frame.

The principal axes of the ellipsoid are those where $\cal{H}$ is
real with $\sigma=0$. A heavy-hole ground state implies
$\Delta_{LH0}>0$, and the ellipsoid is oblate (flat). A light-hole
ground state corresponds to a prolate (elongated) ellipsoid. The
in-plane ellipticity is measured by $\rho_0$. An example relevant
for the present study is given in Fig.~\ref{fig:dip}(a).

\bibliographystyle{apsrev4-1}

%

%
%
%

\end{document}